# Performance Evaluation and Optimization of Math-Similarity Search

Qun Zhang and Abdou Youssef

Department of Computer Science
The George Washington University
Washington DC, 20052, USA

*Abstract*

*Similarity search in math is to find mathematical expressions that are similar to a user's query. We conceptualized the similarity factors between mathematical expressions, and proposed an approach to math similarity search (MSS) by defining metrics based on those similarity factors [11]. Our preliminary implementation indicated the advantage of MSS compared to non-similarity based search. In order to more effectively and efficiently search similar math expressions, MSS is further optimized. This paper focuses on performance evaluation and optimization of MSS. Our results show that the proposed optimization process significantly improved the performance of MSS with respect to both relevance ranking and recall.*

## 1. Introduction

Math is a symbolic language with levels of abstraction, has rich structural information, and features considerable synonymy and polysemy among expressions. These characteristics makes math search especially challenging, and calls for similarity search. In our recent paper [11], we proposed an approach to math similarity search (MSS) for Strict Content MathML encoded math expressions. We identified conceptual factors of math similarity, and deduced a math similarity metric. For a query, MSS was designed to help us find math expressions with taxonomically similar functions, hierarchically similar structure, and/or semantically similar data types. It was found to be able to greatly improve the performance of a math search system.

Ideally, performance evaluation of a search engine requires a benchmark infrastructure. However, unlike the already matured text search area, math search is still evolving. The collective effort of the mathematical community in creating benchmarks and reference data is greatly needed. It is beginning to get attention in some recent initiatives such as the NTCIR [6] math task led by Aizawa and Kohlhase, *et al*.

However, to the best of our knowledge, there is not yet a standard Strict Content MathML encoded math query benchmark infrastructure available to us. To create a full benchmark that can be agreed upon requires significant expertise. However, we do need some ground truth for our performance evaluation and system optimization. So for our research purpose, we leverage the DLMF [2] repository.

The rest of the paper starts with a brief summary of the related work in Section 2. It then presents a recap of the conceptual similarity factors and the similarity metric of MSS [11] in Section 3. Section 4 gives in detail the process of performance evaluation and optimization, and Section 5 presents the performance results. Finally, the conclusions are drawn in Section 6.

## 2. Background

An important aspect of search is relevance ranking. Therefore, comparison between ranking schemes is a significant piece of the search system performance evaluation. The various existing ranking comparison metrics can be classified into two categories: subjective comparison metrics and statistical comparison metrics.

The former is purely based on human judgment. It is good for judging the ranking produced by a specific search tool. Some existing research work includes [4] and [9].

Statistical comparison metrics are often based on correlation measurement, which is an extension of the distance measure between two permutations [1], [3]. These works leverage some classical metrics such as *Kendall*'s $\tau$ [5], and the standard correlation coefficient, also known as *Spearman*'s $\rho$ [8]. This type of ranking comparison metric can be programmatically implemented to save human labor and avoid human bias in the comparison. Thus, the statistical ranking comparison approach is selected for our research.

As mentioned earlier, performance evaluation of search systems demands a benchmark infrastructure. The NTCIR-11 Math-2 task [6] and its optional Wikipedia subtask [7] are the only significant ones for math search evaluation, as far as we know. The target dataset of NTCIR-11 Math-2 consists of about 100,000 scientific articles in ArXiv that are converted from LaTeX to an XHTML format. NTCIR-11 Math-2 provides some sample queries, but there are no expected results as reference for systematic evaluation. Rather, a pooling based relevance judgment is proposed to run manually by human reviewers' subjective assessment.

The NTCIR-11-Math-Wikipedia-Task, on the other hand, uses the English Wikipedia as a test collection which is more suitable for general users compared with the NTCIR-11 Math-2. It also provides a number of sample queries which include 2 content queries that are encoded in Content MathML. However, there are no expected results for participants to do self-evaluation, or any formal evaluation conducted by the organizers except a final judgment based on participants' oral presentations.

Both of the above research initiatives encode their dataset in XHTML, which is not yet converted to Strict Content MathML, while our MSS is for Strict Content MathML encoded math expressions only. This makes it challenging for us to participate in these collective efforts at this moment. Additionally, none of the above tasks provides any reference results for us to use as ground truth to do performance evaluation; instead, the assessment is either done by their organizers in a pooling based manual process or not available at all. This prevents us from obtaining performance evaluation benefits as of now. However, it is anticipated that these resources can be leveraged in the future once the Strict Content MathML encoded dataset and the full set of ground truth with both queries and expected results become available, so that more thorough and more objective performance evaluation can be done.

## 3. Math-Similarity Search (MSS)

The goal of MSS is that, given a math expression that is encoded in Strict Content MathML, to identify a list of structurally and semantically similar math expressions from a library of Strict Content MathML encoded math expressions, and sort the list by similarity according to some similarity measure. In [11], we identified conceptual factors that capture aspects of math similarity, and came up with a similarity metric that takes all those factors into consideration.

### 3.1 Math Similarity Factors

We identified five major factors to math similarity measure, with focus on the structural and semantic aspects of math expression.

(1) Data Type
This factor is captures the superior significance (to similarity) of functions and operators over arguments or operands (for our research purpose, there is no distinction between function and operator), and of structure over notation.

For example, query $F = G\frac{m_1 m_2}{r^2}$, Newton's law of universal gravitation, is more similar to $F = \text{ke}\frac{q_1 q_2}{r^2}$, Coulomb's law, than to a hypothetical expression $F + G + m_1 + \frac{m_2}{r^2}$, containing the same set of operands and notations as the query.

(2) Taxonomic Distance of Functions
The math function's taxonomy information is available to us through the content dictionary (CD) attribute of Strict Content MathML. Functions in the same CD are assumed to be more similar than functions in different CDs.

(3) Match-Depth
The more deeply nested a query is in an expression, the less similarity there is between the query and that expression. Therefore, match-depth is used as a similarity decaying multiplicative factor. One can utilize different models for this decay factor, such as the exponential model, the linear model, the quadratic model, or the logarithmic model.

(4) Query Coverage

How much of the query is covered in the returned expression is important. In general, the greater the query coverage is, the greater the significance is to the similarity measure.

(5) Formula vs. Non-Formula Expression

Typically in math content, formulas carry more weight than non-formula expressions. Furthermore, because equality often implies definition or 100% precision, it should be weighted higher than inequality.

### 3.2 Math Similarity Metric

MSS takes Strict Content MathML parse trees as the primary model representing math expressions. The similarity between two math expressions is defined recursively based on the height of the corresponding parse trees.

When the height is 0, the parse tree is a single node which can represent a constant, a variable, or a function. The similarity between two parse trees, $T_1$ and $T_2$ are defined as:
- If $T_1$ and $T_2$ are constants: $sim(T_1, T_2) = 1$, if $T_1 = T_2$; otherwise, $sim(T_1, T_2) = \delta$, $0 \leq \delta < 1$.
- If $T_1$ and $T_2$ are variables: $sim(T_1, T_2) = 1$, if $T_1 = T_2$; otherwise, $sim(T_1, T_2) = \zeta$, $0 \leq \zeta \leq 1$.
- If $T_1$ and $T_2$ are functions:
    - $sim(T_1, T_2) = 1$, if $T_1$ and $T_2$ are the same function
    - $sim(T_1, T_2) = \mu$, if $T_1$ and $T_2$ are different functions of the same category in the taxonomy, where $0 < \mu < 1$
    - $sim(T_1, T_2) = 0$, if $T_1$ and $T_2$ are different functions of different categories
- If $T_1$ and $T_2$ are not of the same data type:
    - $sim(T_1, T_2) = \theta$, if one is constant and the other is a variable, where $0 \leq \theta < 1$.
    - $sim(T_1, T_2) = 0$, if one is a function and the other is either constant or variable

*Note that $\delta$, $\zeta$, $\mu$, and $\theta$ are parameters that will be optimized experimentally.*

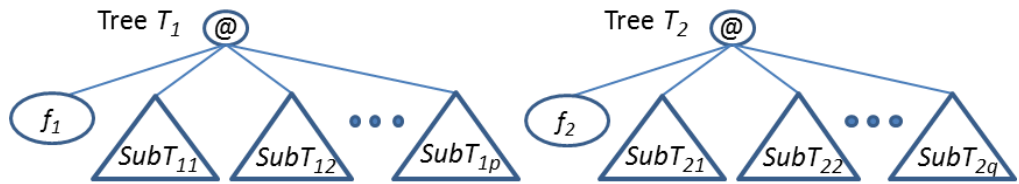

**Figure 1: Illustration of two trees $T_1$ and $T_2$ of height $h \geq 1$**

When the heights of the two trees $T_1$ and $T_2$ are ≥ 1, each of them is composed of function *apply* operator as root, a leftmost child representing the function name, followed by a list of arguments which are sub-trees, as illustrated in Fig. 1. The similarity between $T_1$ and $T_2$ is defined as a weighted sum of the similarity between the two functions $f_1$ and $f_2$, and the similarity between the two lists of arguments:

$sim(T_1, T_2) = \alpha \cdot sim(f_1, f_2) + \beta \cdot sim(\{SubT_{11}, SubT_{12}, \ldots SubT_{1p}\}, \{SubT_{21}, SubT_{22}, \ldots SubT_{2q}\})$,

where $\alpha = \frac{\omega}{p+\omega}$ and $\beta = \frac{1}{p+\omega}$ are weighting factors, $\omega$ ($> 1$) is a boost value (for boosting functions over arguments), and

$0 \leq sim(\{SubT_{11}, SubT_{12}, \ldots SubT_{1p}\}, \{SubT_{21}, SubT_{22}, \ldots SubT_{2q}\}) \leq p$

The similarity measurement between the two lists of arguments depends on whether the two functions, $f_1$ and $f_2$, are commutative. If both are non-commutative functions, the order of the arguments is observed. The similarity between the two lists of arguments is the sum of the similarities between the corresponding available pairs of argument sub-trees with one from each tree:

$sim(\{SubT_{11}, SubT_{12}, \ldots SubT_{1p}\}, \{SubT_{21}, SubT_{22}, \ldots SubT_{2q}\}) = \sum_{i=1}^{min(p,q)} (sim(SubT_{1i}, SubT_{2i}))$

Otherwise, the similarity between the two lists of arguments is the similarity of the best match between the two lists:

$sim(\{SubT_{11}, SubT_{12}, \ldots SubT_{1p}\}, \{SubT_{21}, SubT_{22}, \ldots SubT_{2q}\})$
$= max\{(\sum_{i=1}^{p} (sim(SubT_{1i}, SubT_{2\,t(q, p, i)}))) | t(q, p, i)$ is the i-th element of a p-permutation of $q\}$
$\approx \sum_{i=1}^{min(p,q)} (max\{sim(SubT_{1i}, SubT_{2\,\varphi(i)}) |$ applying greedy approximation, $\varphi(i) = 1, 2, \ldots, q$ and $\varphi(i) \notin \{\varphi(1), \varphi(2), \ldots \varphi(i-1)\}\})$

For partial match consideration, not only the similarity between $T_1$ and $T_2$ at their root level is measured, but also the similarity between entire tree $T_1$ and each single sub-tree of $T_2$, as well as the similarity between entire tree $T_2$ and each single sub-tree of $T_1$ are measured in order to find the best match among all of these possibilities. Assuming $T_1$ is the parse tree for query,

$sim(T_1, T_2) = max\{$
$(\alpha \cdot sim(f_1, f_2) + \beta \cdot sim(\{SubT_{11}, SubT_{12}, \ldots SubT_{1p}\}, \{SubT_{21}, SubT_{22}, \ldots SubT_{2q}\}))$,
$dp \cdot max\{sim(T_1, f_2), sim(T_1, SubT_{21}), sim(T_1, SubT_{22}), \ldots, sim(T_1, SubT_{2q})\}$,
$cp \cdot max\{sim(f_1, T_2), sim(SubT_{11}, T_2), sim(SubT_{12}, T_2), \ldots, sim(SubT_{1p}, T_2)\}\}$,

where $dp$ and $cp$ are the match-depth penalty factors of different values, where dp corresponds to the depth of $T_1$ in $T_2$, and cp corresponds to the depth of $T_2$ in $T_1$.

## 4. Performance Evaluation and Optimization

Even though the preliminary implementation of MSS showed some promising advantages over the non-similarity based search, the many parameters of the MSS similarity metric must be optimized.

### 4.1 Evaluation Methodology

As the DLMF math digital library is among the few available and easily accessible for us, this research leverages the DLMF as the source for mathematical expressions repository, and we compare our similarity search to the DLMF search system. Since there is no Strict Content

MathML encoding of the DLMF, a significant subset of the DLMF is hand-crafted into Strict Content MathML encoding in this research to create the test database. Then queries with varying degrees of mathematical complexity and length were selected and encoded in Strict Content MathML. Some sample queries are listed in [11] for reference. For each query in the test set, we identify the expected relevant expressions from the subset of DLMF source repository used as a test database, and further rank them manually in a relevancy based order by a group of human experts, which are then used as a ground truth.

Each query is compared with the expressions in the test database, and a similarity value is computed with the MSS similarity metric described in Section 3. The full list of the expressions is then sorted by in descending order of similarity. Depending on the number of relevant expressions in the ground truth for this particular query, the top "N" number of expressions is selected to form the MSS hit list for this query.

Up to this point, for any given query, there are three hit lists: one from the ground truth, another from the DLMF site returned by DLMF search, and a third from the similarity search MSS. We compare both the MSS hit list and the DLMF hit list with the ground truth list, based on the performance metric described in Section 4.2. The output of such comparison shows the alignment of a hit list with the ground truth.

### 4.2 Performance Metric

We evaluate the performance with respect to both recall and relevance ranking. The evaluation is done for each sample query, before the final average across all of the sample queries is computed.

The recall measurement is fairly standard. In addition to the recall of all of the results in the ground truth hit list, we also measure the recall of the top 10 results relative to the ground truth

For relevance ranking comparison, we measure the correlation between the ranked list that is returned from each search system, and the ground truth list for each sample query. The correlation is measured by the standard correlation coefficient Spearman's $\rho$, and Kendall's $\tau$. $\rho$ is standard correlation coefficient of statistical dependence between two variables. It focuses more on absolute order, i.e. where each hit is ranked. $\tau$ is to measure the extent of agreement between two lists. It focuses more on relative order of the hits, i.e. which comes before which. These two correlation metrics complement each other in serving as performance indicators.

Following the standard practice for correlation analysis, for both $\rho$ and $\tau$, we refer to their critical value tables to judge statistical significance of the correlation under evaluation. For any given sample query, depending on the size of its ground truth, i.e. the number of similar expressions expected to return, the corresponding critical value is looked up and compared with the computed correlation measure. The correlation appears statistically significant with the specific level of confidence only if the correlation measure reaches or exceeds the critical value of that confidence level. In our research, two different confidence levels are considered: 95% and 99%.

**4.3 Optimization Concerns and Motivation**

In [11] many parameters of the MSS metric need be optimized, such as taxonomic distance values (e.g. µ) between functions, function nodes type booster value ω, query coverage factor, etc. In the early implementation of the MSS metric, random values were used to initialize the parameters. With preliminary experimentation being done, some reasonable values were obtained, so that the MSS showed some performance advantage compared with other non-similarity based math search. This was indicated in [11]. In this paper, we conduct a systematic experimentation based optimization to arrive at optimal values of all the parameters.

One of the similarity factors, the match-depth, is represented as a similarity-decaying multiplicative factor in our MSS metric. As mentioned earlier, there can be several alternative models for such decay, such as exponential decay, linear decay, quadratic decay, or logarithmic decay. Which of these decay models performs better for MSS metric, and how to configure these decay models along with other parameters of the MSS metric, become part of the concerns that motivate us to start the optimization process.

In the preliminary evaluation conducted in [11], the number of results in the ground truth for each sample query was normalized to 20. However, in reality the actual number of expressions in the test database that are similar to each query may not be the same – if less than 20, then the ground truth was padded with "weeds"; and if larger than 20, then normalizing the ground truth size to 20 would limit the view or window of expressions to examine. Either case can impact the overall system performance, e.g. recall. For this reason, we choose to remove the ground truth size normalization across all the queries, and leave the ground truth of all the sample queries as is.

**4.4 The Optimization Process**

Motivated by the above optimization concerns, we conduct a systematic optimization process to more effectively and efficiently optimize MSS. The whole optimization process is illustrated in Fig. 2.

This optimization process includes multiple generations of optimization. Each generation has its own optimal trial value for each of the parameters, and its own best model for the match-depth factor. Within one generation, all the parameters are to be optimized. For each parameter, we step through different trial values with equal increments. For each selected trial value, we measure the MSS system performance for each sample query by using the performance metric to measure the recall and correlation between the MSS hit list and the ground truth list for that query. After that, an average of those recall and correlation measurements from each query execution is obtained, and referenced for the selection of best trial value for that specific parameter.

```
begin
    for each different match-depth model
        for each generation of parameters do
            for each parameter
                for trial-value of current parameter
                    for each query
                        run MSS search;
                        collect similarity measurement;
                        get hits;
                        conduct recall & correlation analysis;
                    end-for
                    evaluate average performance across all queries;
                end-for
                pick the best trial-value for current parameter
            end-for
            if (this generation did not improve over previous generation)
                stop generation evolution;
                flag "converged" for current model;
            end-if
        end-for
    end-for
    pick the best model set;
end
```

Figure 2: Illustration of MSS Optimization Process

How to decide when to stop the generational evolution, needs a "convergence" criterion, i.e. a measurement of the overall system performance across all of the queries for a given system configuration. This is similar to the best trial value selection for a specific parameter within a generation. Once such overall system performance of a generation does not show any improvement compared to the previous generation, the generational evolution terminates for the given match-depth model. However, the full optimization process does not stop until the above generational evolution is done for each the different match-depth models, and then the best model is identified at the end.

## 5. Optimization Results

In this section we show the findings of the optimization described in the previous section. Figure 3 shows the performance data of the optimized MSS with the exponential match-depth model: $a^k$, where $k$ is the match-depth. The top two charts show the relevance ranking over the sample queries where the left one shows the $\tau$ metric, and the right one shows the $\rho$ metric. Both of these two charts have reference critical values marked out, where the green solid line indicates the critical values at 95% confidence level, and the purple dotted line indicates the critical values at 99% confidence level. The fact that the MSS's correlation measures are mostly above the critical values indicate the improvement over DLMF search is statistically significant. The bottom two charts show the recall over the sample queries, where the left shows the overall recall, and the right one shows the recall relative to the top ten results. Both recall and correlation show the advantage of MSS over DLMF search, and show substantial improvement in performance throughput over the preliminary implementation presented in [11]. Overall recall improves 76% reaching 0.7467, and recall of top 10 hits improves 51% reaching 0.8575.

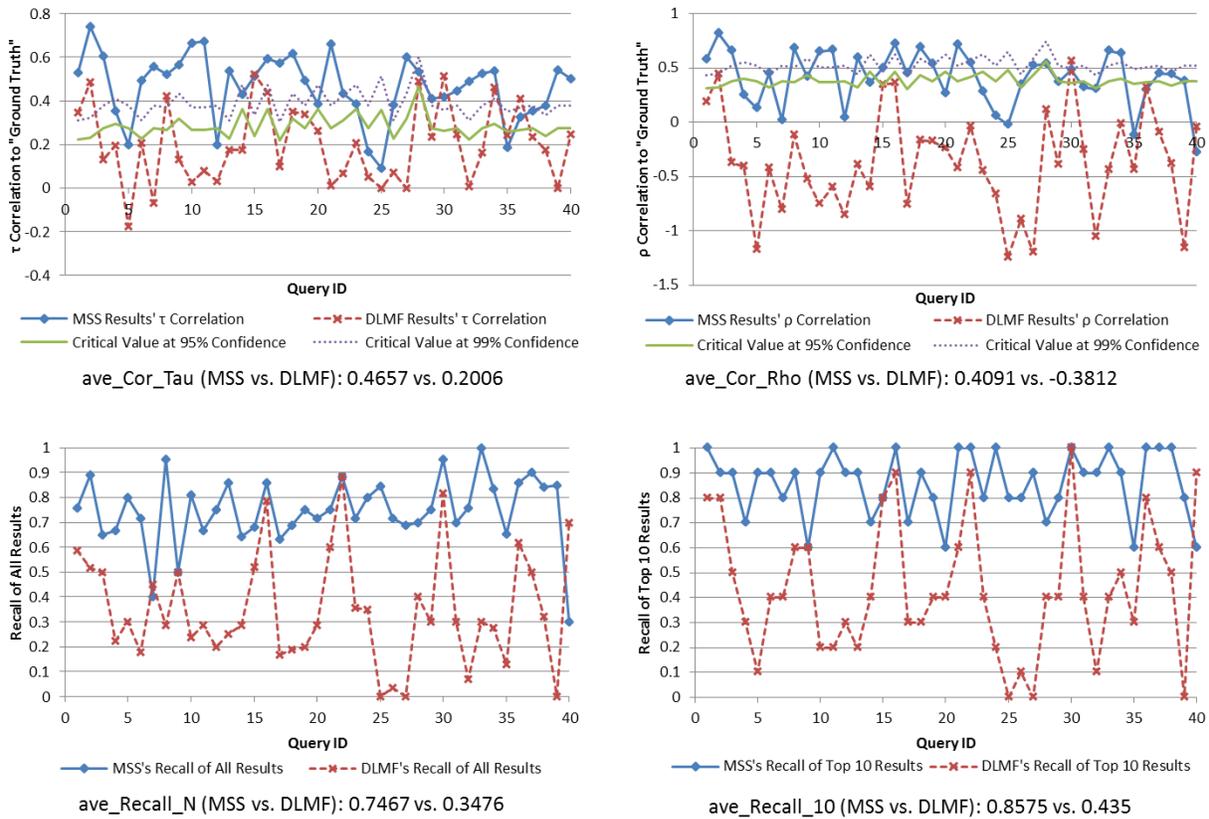

**Figure 3: Performance Data of Optimized MSS with Exponential Match-Depth**

Figure 4 shows the performance data of the optimized MSS with the linear match-depth model: $\max(1 - b \cdot k, \ \varepsilon)$, where $k$ is the match-depth and $\varepsilon$ is a small positive constant to ensure that the value of the decay factor stays positive. The linear match-depth model based MSS at its optimal setting outperforms the DLMF search with respect to both correlation and recall. Figure 5 shows the performance data of the optimized MSS with the quadratic match-depth model: $\max(1 - c \cdot k^2, \ \varepsilon)$, where $k$ and $\varepsilon$ are defined as those in linear model. The quadratic match-depth model based MSS however underperforms the DLMF search in correlation and recall of the top ten results, even though the overall recall at its optimal setting is slightly higher than that of DLMF search. Based on our experimentations, the performance of both of these two models cannot reach that of the exponential model.

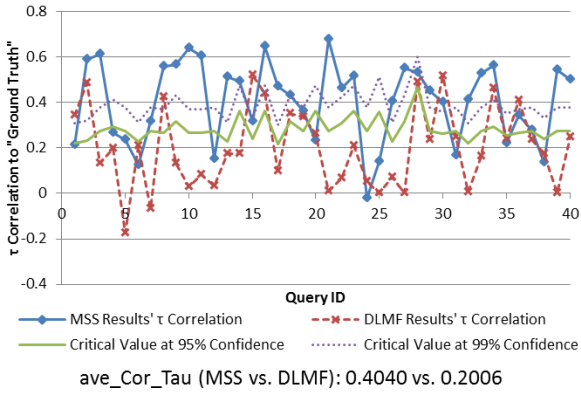
ave_Cor_Tau (MSS vs. DLMF): 0.4040 vs. 0.2006

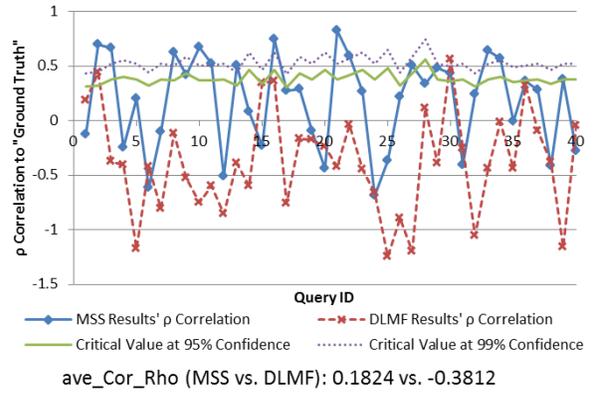
ave_Cor_Rho (MSS vs. DLMF): 0.1824 vs. -0.3812

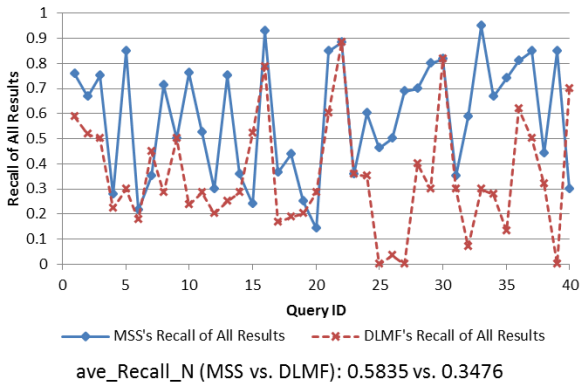
ave_Recall_N (MSS vs. DLMF): 0.5835 vs. 0.3476

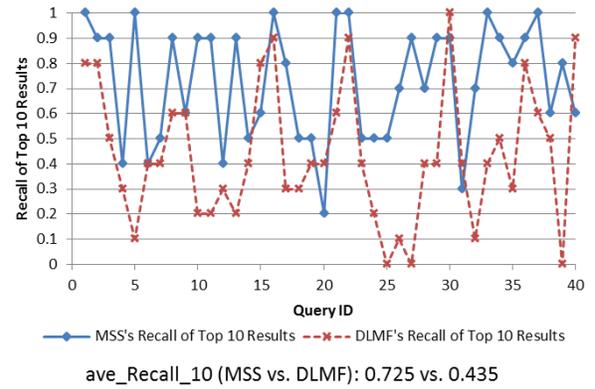
ave_Recall_10 (MSS vs. DLMF): 0.725 vs. 0.435

**Figure 4: Performance Data of Optimized MSS with Linear Match-Depth**

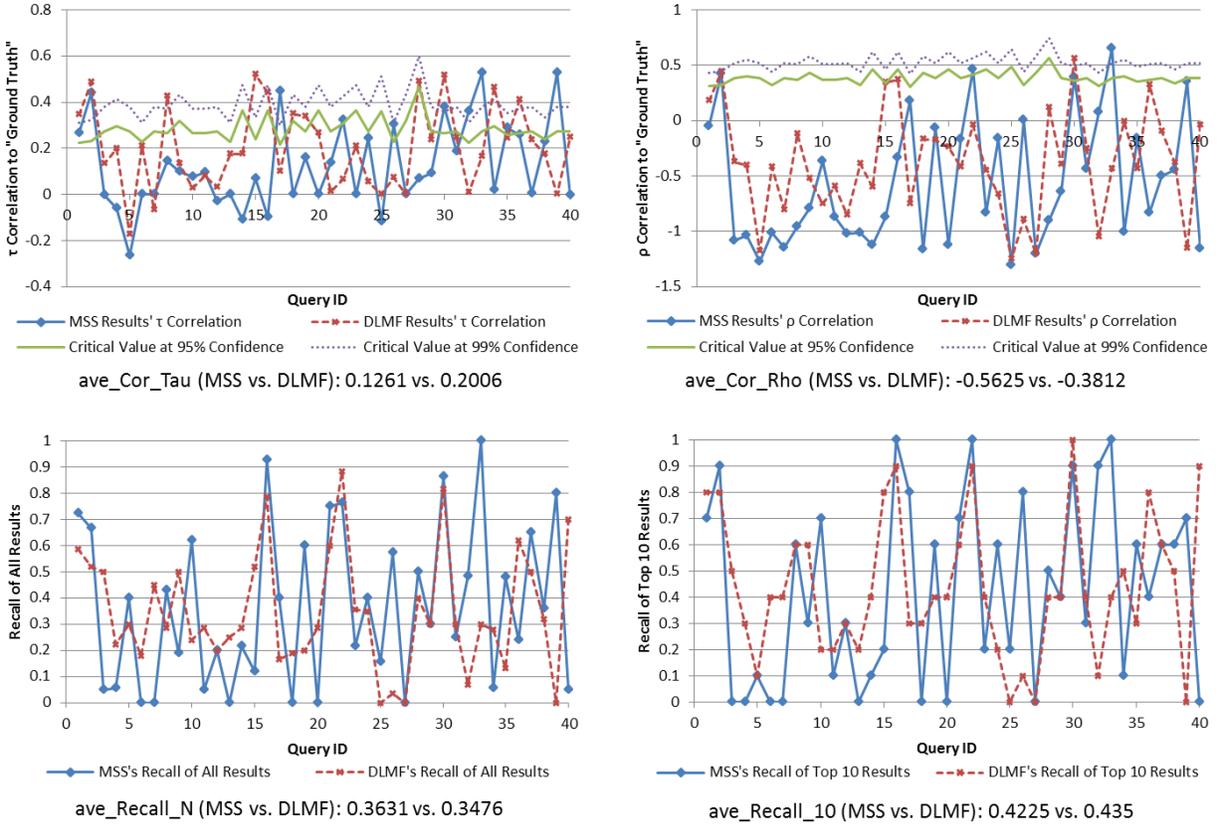

**Figure 5: Performance Data of Optimized MSS with Quadratic Match-Depth**

Figure 6 shows the performance data of the optimized MSS with the logarithmic match-depth model: $\max(1 - d \cdot \ln(k + 1),\ \varepsilon)$, where $k$ is the match-depth and $\varepsilon$ are as before. Interestingly, as shown in the performance data charts of Fig. 6, our experimentation indicates that the logarithmic match-depth model performs the best – the recall and correlation measurements at the optimal setting are all higher than those of other match-depth models. The $\tau$ and $\rho$ correlation measures reach the greatest: 0.4737 and 0.4928 respectively, which are 136% and 207% higher than those of DLMF search, while the overall recall and the recall of top ten results reach 0.8251 and 0.9075, which are 137% and 109% higher than those of DLMF search respectively. Clearly, MSS with the logarithmic match-depth model shows not only significantly greater relevance ranking and recall, but also more consistency. The logarithmic match-depth model demonstrates the maximal advantage of MSS over DLMF search.

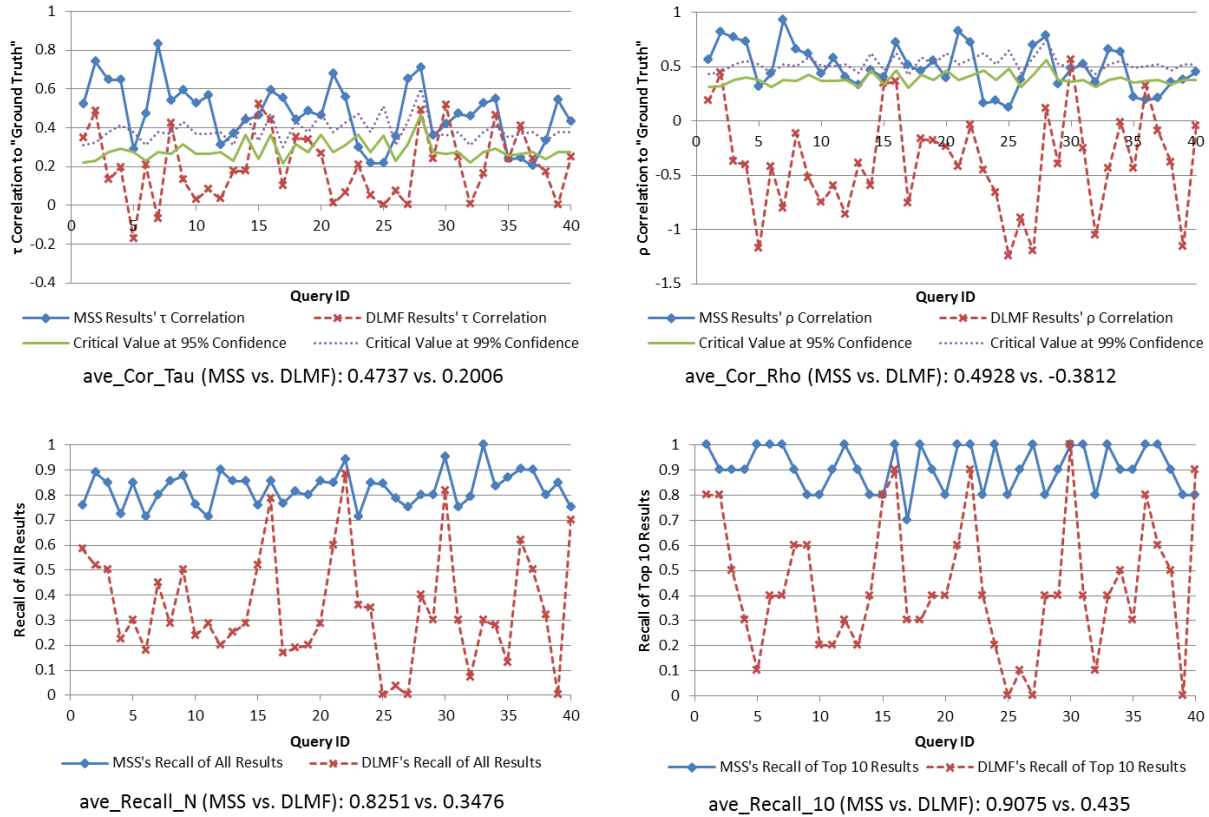

**Figure 6: Performance Data of Optimized MSS with Logarithmic Match-Depth**

The MSS with the logarithmic match-depth model shows its optimal performance after ten generations of optimization. Figure 7 shows the relevance ranking improvement throughout the optimization process, in terms of the $\tau$ and $\rho$ correlation measures. Figure 8 shows the recall improvement, with respect to both the overall recall and the top-10 recall. Both of these figures also show the performance data of DLMF search, which appears constant across the different generations of optimization of MSS. As shown in both Fig. 7 and Fig. 8, the performance improvement over the evolution of the initial five generations is more significant than that of the later five generations. Over the complete ten generations of optimization process, all of the four performance metrics data improved substantially with more than 20% growth, among which the $\rho$ correlation measure more than doubled.

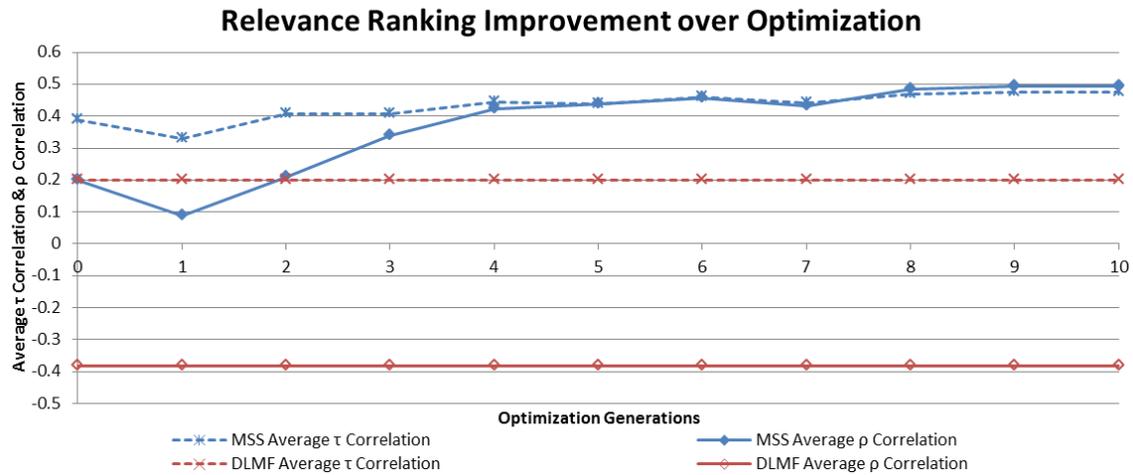

Figure 7: Relevance Ranking Improvement over Optimization of

MSS with Logarithmic Match-Depth Model

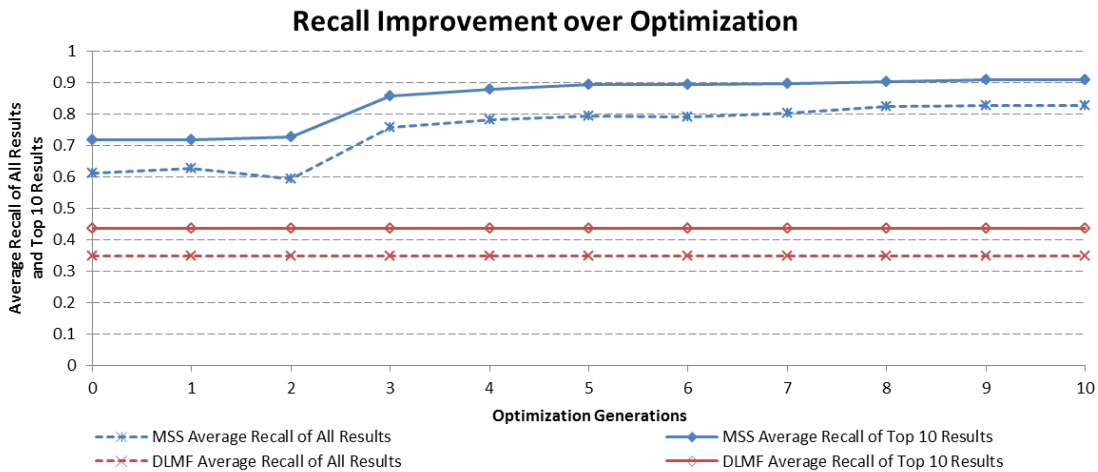

Figure 8: Recall Improvement over Optimization of MSS with Logarithmic Match-Depth Model

In order to avoid the problem of over-fitting of the parameters to the training data, and to obtain a more objective assessment of performance, we conducted cross-validation-based optimization and performance evaluation.

Generally in cross-validation, we train (optimize) a system on a dataset called the "training set", and perform testing on a different dataset called the "validation set" or "test set". To adopt the cross-validation technique in optimization and performance evaluation of MSS system, we partition all of the 40 sample queries randomly into two disjoint subsets of equal size, one of which is used as the training set for optimization and the other as test set for performance evaluation. Such crossvalidation is applied to MSS system with all the four different match-depth models. The performance results are shown in Table 1.

As shown in Table 1, with cross-validation, the performance results of MSS largely agree with those obtained without cross-validation. Particularly, both with and without cross-validation, it turns out that the logarithmic match-depth decaying model is the best model of all the four alternatives, and under that model, all of our four performance metrics achieve the same level with as without cross-validation. This lends more confidence to the robustness of the optimized values of the parameters, and more validation to our results.

Table 1: Performance Result Without vs. With Cross Validation

| Math Search System | | Recall Metric | | Relevance Ranking Metric | |
|---|---|---|---|---|---|
| | | Ave. Overall Recall | Ave. Top 10 Recall | Ave. ρ Correlation | Ave. τ Correlation |
| MSS (Exponential Model) | W/O Cross-Validation | 0.7467 | 0.8575 | 0.4091 | 0.4657 |
| | **W/ Cross-Validation** | **0.7785** | **0.87** | **0.4104** | **0.4490** |
| MSS (Linear Model) | W/O Cross-Validation | 0.5835 | 0.725 | 0.1824 | 0.4040 |
| | **W/ Cross-Validation** | **0.5904** | **0.735** | **0.1943** | **0.4168** |
| MSS (Quadratic Model) | W/O Cross-Validation | 0.3631 | 0.4225 | -0.5625 | 0.1261 |
| | **W/ Cross-Validation** | **0.2796** | **0.27** | **-0.7955** | **0.0654** |
| MSS (Logarithmic Model) | W/O Cross-Validation | 0.8251 | 0.9075 | 0.4928 | 0.4737 |
| | **W/ Cross-Validation** | **0.8288** | **0.89** | **0.4807** | **0.4528** |
| DLMF | on 40 Full Set | 0.3476 | 0.435 | -0.3812 | 0.2006 |
| | **on 20 Test Set** | **0.3729** | **0.47** | **-0.3561** | **0.2117** |

## 6. Summary and Conclusions

MSS takes structural aspects of math expressions into serious consideration, honors the semantic significance of function as opposed to argument, and further differentiates functions by their taxonomic distance. With MSS, partial match of query can be detected and weighted based on the coverage. Additionally, equations (a.k.a. formulas) are differentiated from expressions (non-formulas) in MSS, and ranked higher. All of these make MSS discover more effectively the math expressions that are similar to queries, and rank them better.

In order to quantitatively evaluate the performance of MSS, relevance ranking and recall based performance metrics were used. Furthermore, a systematic optimization process was conducted which improved the performance of MSS substantially. Our experiments show that the MSS with a logarithmic match-depth decay model performs better than the MSS with other match-depth decay models. Finally, cross-validation was applied to the optimization and performance evaluation, and obtained the results lead that agreed with the conclusions drawn through the prior experiments. The performance data after optimization validate more sufficiently the advantage of math similarity search over non-similarity based search typified by the DLMF.

## References

[1] Judit Bar-Ilan. Comparing rankings of search results on the Web. *Information Processing & Management, 41, 1511-1519, Elsevier, 2005*.

[2] The Digital Library of Mathematical Functions (DLMF), the National Institute of Standards and Technology (NIST). http://dlmf.nist.gov/

[3] Ronald Fagin, Ravi Kumar, and D. Sivakumar. Comparing top k lists. *ACM-SIAM Journal on Discrete Mathematics, 17(1), 134-160, 2003*.

[4] D. Hawking, N. Craswell, P. Bailey, and K. Griffiths. Measuring search engine quality. *Information Retrieval, 4, 33-59, Springer Netherlands, 2001*.

[5] Maurice G. Kendall. Rank Correlation Methods. *New York, Hafner Publishing Co., 1955*.

[6] The 11[th] National Institute of Informatics Testbeds and Community for Information access Research Workshop. http://ntcir-math.nii.ac.jp/. *2013 - 2014.*

[7] An optional free subtask of the NTCIR-11 Math-2 Task. http://ntcir11-wmc.nii.ac.jp. *2014.*

[8] Charles Spearman. The proof and measurement of association between two things. *Amer. J. Psychol. 15: 72–101*, 1904.

[9] L. Vaughan. New measurements for search engine evaluation proposed and tested. *Information Processing & Management, 40(4), 677-691, Elsevier, 2004*.

[10] Abdou Youssef. Methods of Relevance Ranking and Hit-Content Generation in Math Search. *Calculemus/MKM, 2007*.

[11] Qun Zhang and Abdou Youssef. An Approach to Math-Similarity Search. *International Conference, CICM 2014, Coimbra, Portugal, July 7-11, 2014. Proceedings pp 404-418. 2014.*